\documentclass{tech-report}
\pdfoutput=1

\title{A Berkeley View of Teaching CS at Scale}
\author{Kevin Lin}
\newcommand*{\researchadvisor}{John S. DeNero}
\newcommand*{\secondreader}{Joshua A. Hug}

\newcommand*{\signed}{}

\ifdefined\signed
    \usepackage{pdfpages}
\fi

\bibliography{bibliography}

\begin{document}

\ifdefined\final
    \setcounter{page}{3} 
\else\ifdefined\signed
    \includepdf{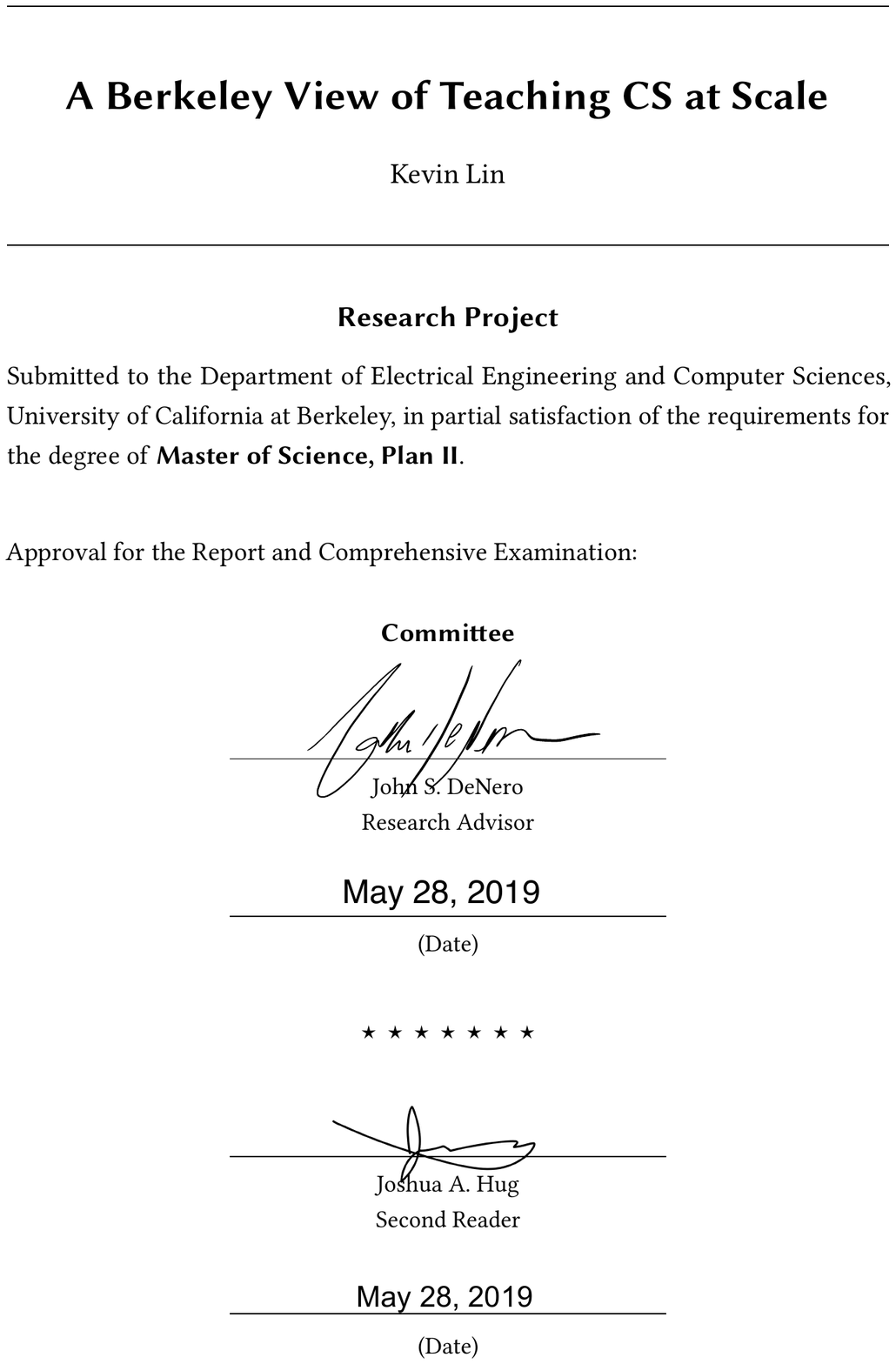}
    \clearpage
\thispagestyle{empty}
\vspace*{3in}

\begin{center}
Copyright \copyright{} \the\year, by the author(s).

All rights reserved.
\end{center}

\noindent
Permission to make digital or hard copies of all or part of this work for personal or classroom use is granted without fee provided that copies are not made or distributed for profit or commercial advantage and that copies bear this notice and the full citation on the first page. To copy otherwise, to republish, to post on servers or to redistribute to lists, requires prior specific permission.
\else
    \newlength{\rulewidth}
\setlength{\rulewidth}{2.5in}
\newlength{\ruleheight}
\setlength{\ruleheight}{0.5pt}

\begin{titlepage}
\begin{addmargin}{0.45in}

\begin{center}
    \rule{\linewidth}{\ruleheight}
    \vspace*{2ex}

    \makeatletter
    \begin{LARGE}
        \textbf{\@title}
    \end{LARGE}
    \\ [3ex]
    \begin{large}
        \@author
    \end{large}
    \makeatother

    \vspace*{2ex}
    \rule{\linewidth}{\ruleheight}
\end{center}

\vspace*{2ex}
\begin{center}
    \begin{large}
        \textbf{Research Project}
    \end{large}
\end{center}

\noindent
Submitted to the Department of Electrical Engineering and Computer Sciences, University of California at Berkeley, in partial satisfaction of the requirements for the degree of \textbf{Master of Science, Plan II}.

\vspace*{5ex}
\noindent
Approval for the Report and Comprehensive Examination:

\vspace*{2ex}
\begin{center}
    \textbf{Committee}

    \begin{small}
        \vspace*{8ex}
        \rule{\rulewidth}{\ruleheight} \\
        \researchadvisor \\
        Research Advisor

        \vspace*{5ex}
        \rule{\rulewidth}{\ruleheight} \\
        (Date)

        \vspace*{5ex}
        \(\star\; \star\; \star\; \star\; \star\; \star\; \star\)

        \vspace*{8ex}
        \rule{\rulewidth}{\ruleheight} \\
        \secondreader \\
        Second Reader

        \vspace*{5ex}
        \rule{\rulewidth}{\ruleheight} \\
        (Date)
    \end{small}
\end{center}

\end{addmargin}
\end{titlepage}
    \clearpage
\thispagestyle{empty}
\vspace*{3in}

\begin{center}
Copyright \copyright{} \the\year, by the author(s).

All rights reserved.
\end{center}

\noindent
Permission to make digital or hard copies of all or part of this work for personal or classroom use is granted without fee provided that copies are not made or distributed for profit or commercial advantage and that copies bear this notice and the full citation on the first page. To copy otherwise, to republish, to post on servers or to redistribute to lists, requires prior specific permission.
\fi\fi

\begin{abstract}
    \noindent
    Over the past decade, undergraduate Computer Science (CS) programs across the nation have experienced an explosive growth in enrollment as computational skills have proven increasingly important across many domains and in the workforce at large. Motivated by this unprecedented student demand, the CS program at the University of California, Berkeley has tripled the size of its graduating class in five years. The first two introductory courses for majors, each taught by one faculty instructor and several hundred student teachers, combine to serve nearly 2,900 students per term. This report presents three strategies that have enabled the effective teaching, delivery, and management of large-scale CS courses: (1) the development of autograder infrastructure and online platforms to provide instant feedback with minimal instructor intervention and deliver the course at scale; (2) the expansion of academic and social student support networks resulting from changes in teaching assistant responsibilities and the development of several near-peer mentoring communities; and (3) the expansion of undergraduate teacher preparation programs to meet the increased demand for qualified student teachers. These interventions have helped both introductory and advanced courses address capacity challenges and expand enrollments while receiving among the highest student evaluations of teaching in department history. Implications for inclusivity and diversity are discussed.
\end{abstract}
\tableofcontents

\chapter{Introduction}
\label{chapter:introduction}

\begin{quote}
    Computer science classrooms are overflowing at colleges and universities across the United States. Enrollments are rising quickly, not only for majors, but also for non-majors who recognize the importance of computing skills in today's economy. This enrollment growth puts enormous pressure on computer science departments, which have not been able to expand to keep pace. \cite{Roberts:2018}
\end{quote}
In the decade between 2008 and 2018, CS departments across the United States have experienced double-digit undergraduate enrollment increases while ``the overall growth in teaching capacity woefully lags the growth in students [with] the vast majority of departments [reporting] increased difficulty in managing the situation'' \cite{TaulbeeSurvey2018}. From 2006 to 2015, the average number of CS majors in large departments (25 or more tenure-track faculty) increased from 341 to 970 and for small departments from 158 to 499 majors \cite{GenerationCS}. While growth varies between programs, the data make it clear that ``significant growth is under way at many institutions,'' and that ``the conditions exist for continued growth in the demand for CS and related jobs, degrees, and courses'' \cite{CSUndergraduateEnrollments}.

In this report, we present three strategies which have enabled the effective teaching of large enrollment CS courses at the University of California, Berkeley (UC Berkeley):
\begin{description}
    \item[Automation] The autograder infrastructure and online platforms which are now able to provide instant feedback, minimize manual grading, and deliver courses at scale.
    \item[Support] The expansion of student support networks through changes in teaching assistant responsibilities and the development of several near-peer mentoring communities.
    \item[Preparation] The expansion of undergraduate teacher preparation programs to meet the increased demand for student teachers.
\end{description}
These strategies support recommendations previously published by the Association for Computing Machinery \cite{RetentionCS}, Computing Research Association (CRA) \cite{GenerationCS}, the National Academies \cite{CSUndergraduateEnrollments}, and other research universities \cite{Maher:2015, Malan:2010, Porter:2013, Guo:2013, Hug:2015, Hug:2017, Reges:1988, Roberts:1995, Alvarado:2017, Minnes:2018, Kay:1998}. In light of the national CS capacity crisis and the increasing size of CS courses, this report identifies automation as the force that has driven subsequent changes in support and teacher preparation practices.

\section{National CS Capacity Crisis}

\begin{quote}
    Current pressures on computer science units are extremely difficult to manage and will also intensify if enrollments continue to grow. Institutional administrators need to work with computer science units to find sustainable approaches to meet the student demand, accounting for important factors such as (1) lack of space for classes and units, (2) academic support required, (3) the limited pool of qualified teaching faculty, (4) the goals and needs of nonmajors taking CS classes, (5) the effect of class size on the course experience, and (6) the desired retention of both students and faculty. \cite{GenerationCS}
\end{quote}
According to the 2017 CRA Enrollment Survey, ``66\% of the 134 responding doctoral-granting units reported that the enrollment growth is having a big impact (i.e., causing significant challenges) on their unit,'' with more than 50\% of doctoral-granting institutions citing 6 significantly increasing problems due to growing enrollments: classroom space shortages, insufficient numbers of faculty/instructors, insufficient numbers of teaching assistants (TAs), increased faculty workloads, office space shortages, and lab space shortages \cite{GenerationCS}.

In response to these challenges, more than 50\% of doctoral-granting institutions have already taken 4 actions to manage student enrollments: significantly increase class sizes, increase the number of academic year sections, increase summer offerings, and reduce low-enrollment classes. More than 65\% of doctoral-granting institutions have already taken 4 actions to increase teaching capacity: use undergraduate TAs and tutors, use more adjuncts or visitors as instructors, use graduate students as instructors, and increase the number of teaching faculty \cite{GenerationCS}. A survey of 78 CS professors from 65 different institutions identified the following three most common approaches for addressing the capacity crisis: (1) altering course offerings by increasing class sizes, offering more sections, and reducing elective offerings; (2) hiring more faculty and TAs; and (3) restricting access to classes, directing non-majors to other classes, and ``weeding out'' students \cite{Patitsas:2016}.

Research suggests that certain interventions can significantly affect the recruitment and retention of women and underrepesented minorities (URM) in CS \cite{Cohoon:2002, Babes-Vroman:2017, Rheingans:2018, Newhall:2014, Narayanan:2018, Lewis:2017, GenerationCS, RetentionCS, CSUndergraduateEnrollments, DiversityGapsCS}.
\begin{quote}
    The underrepresentation of women and people from groups underrepresented in computing raises concerns for a variety of reasons, including (1) issues of equity and fairness, (2) the economic and competitive imperative of ensuring a large and diverse U.S. workforce, (3) the fact that better solutions are developed by teams with a diversity of people and perspectives, and (4) the increasing interdependency between American democracy and the ability to understand and navigate the presentation of information through technology. \cite{RetentionCS}
\end{quote}
Course-level and department-level policies can directly affect which students pursue the major or have access to advanced coursework. More than 40\% of doctoral-granting institutions limit enrollments in high-demand courses, advise less-successful students to leave the major, and require that students are in a major or minor in order to enroll in an advanced course \cite{GenerationCS}. Even nominally objective policies such as restricting access to the major based on GPA can disproportionately disadvantage URM students \cite{DiversityGapsCS, RetentionCS}. Furthermore, ``imposing such restrictions makes the relationship between faculty and students adversarial, causing students to become more competitive and, in many cases, angry,'' with students concluding that they aren't wanted and perpetuating the idea that ``computer science [is] competitive and unwelcoming'' \cite{Roberts:2016, Patitsas:2014, Patitsas:2016}.

\begin{quote}
    In the face of increasing enrollments institutions would do well to take lessons from the past. The share of CIS [Computer and Information Science] and CS bachelor's degrees going to women decreased precipitously beginning in the mid-1980s, and again during the dot-com bust. These drops coincided with past peaks in CS degree production, suggesting that high-enrollment conditions or the actions taken by institutions in response to these surges may have contributed to the decrease in representation of women in undergraduate CS during these times. \cite{CSUndergraduateEnrollments}
\end{quote}
Indeed, many of the actions taken by universities today mirror the actions undertaken in the earlier enrollment surge in the 1980s, which included (1) increasing teaching loads and class sizes, (2) hiring more part-time and adjunct faculty, (3) retraining faculty from other disciplines, and (4) limiting enrollments and access to the major \cite{Curtis:1982}. Many departments have since adopted some of the recommendations cited in the 1982 report including diversifying academic opportunities by creating teaching-track faculty positions and using technology to make education more efficient.

CS education has only recently advanced to the national agenda \cite{StateofCS2018}, slowing the adoption of these ideas and practices. ``There are few researchers with CS education PhDs, and right now few or no active formal CS education PhD programs,'' \cite{CSforAll2018} stymieing the development of pedagogical methods and computer science education as a discipline. ``Teaching large computer science courses has become a more specialized endeavor,'' which grows capacity in impacted lower-division courses but results in an increase in student demand for upper-division courses without necessarily solving the underlying instructional bottleneck \cite{Roberts:2016}. There are simply not enough CS teachers. Furthermore, this capacity crisis is occurring at a time of institutional disinvestment due in part to ``administrators who are convinced that they [\dots\unkern] know when students will next lose interest'' but whose ``very decision ensures a capacity collapse'' \cite{Roberts:2016} in spite of evidence pointing to the opposite: ``While there will probably be fluctuations in the demand for CS courses, demand is likely to continue to grow or remain high over the long term'' \cite{CSUndergraduateEnrollments}.

\section{UC Berkeley Case Study}

This report presents a case study of three strategies for teaching CS at scale as developed in the Department of Electrical Engineering and Computer Sciences (EECS) at the University of California, Berkeley. Some historical context is necessary to understand the undergraduate CS education program that allowed the development of these strategies. The effectiveness of implementing them at other institutions will vary based on factors such as the institution's size and values \cite{CSUndergraduateEnrollments}.

There are two paths into the CS program at UC Berkeley:
\begin{description}
\item[EECS] The Electrical Engineering and Computer Sciences major in the College of Engineering, to which students apply directly in their application to the university, with a cohort of about 400 students matriculating in 2018.
\item[LSCS] The Computer Science major in the College of Letters and Sciences, where students are admitted into the college without declaring a major. Letters and Sciences students can declare the CS major after meeting the requirements for the major. During periods of high student demand and low supply, the LSCS declaration process can be very selective \cite{Roberts:2016, Alivisatos:2017}.
\end{description}
In 2019, the LSCS major is a capped major, admitting any student with an average 3.3 GPA across three introductory courses with an appeal process for students near the threshold. In 2018, about 800 students were accepted into the LSCS major. Based on current introductory CS course GPA trends, on expectation, about half of students who take the required courses will be eligible to declare the LSCS major, though these enrollments also include EECS majors and a large number of non-majors. Tracking students by their interest in the LSCS major, between 60--70\% of interested students successfully declare the LSCS major each year. However, this leaves an estimated 470 interested students unable to declare the LSCS major, including 170 women. This also leaves out students who do not even consider CS due to its reputation as a selective, capped major.

In 2018, between EECS and declared LSCS majors, the undergraduate CS program included over 3,200 majors, representing over 10\% of the university's undergraduate student population. In recent years, enrollment pressure has increased not only due to a growth in the number of majors, but also a growth in the number of courses students take per semester. The average number of upper-division EECS courses taken by a CS major throughout their undergraduate degree has recently increased from 5 courses to 7 courses. Students are taking more EECS courses to fulfill major requirements rather than electives offered by other departments. At the same time, the average time to graduation is only 7.89 semesters, as more students in the program are completing their degree in 3 or 3.5 years. Taken together, CS majors are choosing to take more upper-division CS courses in a shorter period of time, inflating enrollment pressure and demand for courses.

Several factors contribute to this growth. Most upper-division CS courses have a short prerequisite chain, usually only requiring the introductory CS sequence, so students can easily switch into another upper-division CS course if enrollment in their first-choice course is full. Furthermore, the program does not require a capstone project that, at many other institutions, introduces an individual advising requirement upon the faculty and consumes student attention in their final year of study. Department surveys show that students' post-graduation plans are increasingly focused on working in software engineering roles, so students value the technical expertise gained from taking technical, CS courses over breadth or personal interest courses.

Demand from non-majors has also increased. Despite the fact that enrollment preference is given to students in the CS program, an increasing number of non-majors are enrolling in upper-division CS courses. Two of the most popular courses (Introduction to Artificial Intelligence; Efficient Algorithms and Intractable Problems) enrolled about 25\% non-majors in 2018. Adjacent major programs including Data Science; Cognitive Science; Applied Mathematics; Engineering Mathematics and Statistics; Statistics; and Industrial Engineering and Operations Research either explicitly require or credit certain upper-division CS courses towards their undergraduate major degrees. This increase in non-major interest in CS courses mirrors the broader, national trend.

It is this context of external and internal demand for computer science that foreshadowed the Data Science undergraduate program, the ``fastest growing program in the history of Berkeley,'' \cite{Alivisatos:2017}.
\begin{quote}
    Berkeley's Data Science education program aims at a comprehensive curriculum built from the entry level upward to meet students' varied needs for data fluency. It includes a diverse constellation of connector courses that allow students to explore real-world issues related to their areas of interest and continues with intermediate and advanced courses that enable them to apply more complex concepts and approaches.\footnote{\url{https://data.berkeley.edu/education}}
\end{quote}
The Division of Data Sciences connects the School of Information, the EECS Department, the Statistics Department, the Berkeley Institute for Data Science (BIDS), and faculty, staff, and students from across campus. Introductory data science courses have been developed with lessons learned from introductory computer science \cite{Swamy:2018}, and upper-division courses are also designed to scale. Starting as a pilot course with 100 students in Fall 2015, enrollment in the introductory data science course, The Foundations of Data Science, reached 1,500 students in Spring 2019, exceeding enrollments in introductory CS that semester. Core data science courses are commonly co-taught by Statistics and Computer Science faculty while connector courses are offered by many departments across campus to meet the diversity in demand for computational literacy and data skills.

\subsection{Course Format}

The typical introductory computer science course uses the following model:
\begin{description}
\item[Lecture] 3 hours per week introducing concepts to the entire class, led by the instructor.
\item[Lab] 1--2 hours per week of hands-on exploration activities, led by a TA, with around 30 students.
\item[Discussion] 1--2 hours per week of group problem-solving, led by a TA, with around 30 students.
\item[Office Hours] A drop-in space for students to ask questions and get help with course concepts and assignments, normally offered on a regular basis by the instructors and TAs. 
\end{description}
The typical upper-level computer science course consists of 3 hours of lecture and 1 hour of discussion section per week. The EECS Department has experimented with other course formats as well. Data Structures and Programming Methodology is offered during summer session in lab-centric instruction format that consists of 1 hour of lecture and 6 hours of lab per week \cite{Titterton:2010}.

In part due to a shortage of large lecture halls, almost all CS courses have begun webcasting lecture. The campus information technology group records live lecture and posts the video online a few hours afterwards. Many students prefer webcasts over live lecture as they can speed-up, slow-down, pause, and rewind the video, so live lecture attendance in large courses rarely exceeds one third of the true class enrollment by the middle of the semester. Furthermore, lab and discussion section attendance is often not mandatory. Students are typically encouraged to participate, and the course policies may provide incentives for attendance, but attendance is rarely required.

As a consequence, there is a significant number of students enrolled in CS courses who rarely attend lecture but still learn all of the course content by watching webcasts. In 2018, this format was officially adopted by Introduction to Database Systems, which was offered to matriculating UC Berkeley students with the regular lecture sessions replaced entirely by professional recordings on the same content by the professor. To keep students on track, the course expects them to submit short, weekly quizzes on basic lecture concepts. Students are encouraged to attend discussion sections and office hours to clarify concepts from the webcast, build problem-solving skills, and collaborate with other students in the course. While relying on online resources frees demand for resources such as seating in lecture, these students often utilize other components of the course such as discussion, lab, office hours, study groups, and tutoring where learning occurs in smaller group environments. Scaling capacity in these activities has become an increasingly important focus for the department, a theme that is revisited throughout this report.
\chapter{Automation}
\label{chapter:automation}

For decades, CS courses at UC Berkeley have used command-line interface autograding utilities such as the \texttt{grading} package. This package allows students to submit files (e.g. programming assignments) to the instructor's UNIX account for grading. Autograding is available through makefiles defined by the instructor for each assignment, and feedback can be automatically emailed to students once the autograder finishes execution. More recent autograding solutions utilize container technology to improve reliability and web frontends to improve the user experience.

In large CS courses, much of the feedback on program correctness is provided by autograders. This has both practical and pedagogical benefits as it reduces the workload of grading student work while supporting students to make progress through independent debugging. These autograding solutions have traditionally been supported by a single computing server provided by the department dedicated to supporting the grading needs of each course. However, when hundreds of students request autograding resources (often in the hours before assignment deadlines), student submissions are placed in a long grading queue. This is an especially important consideration because many UC Berkeley CS courses rely on automated feedback to provide assistance to a large number of students, so a long grading queue is an educational denial of service.

Furthermore, large courses often require additional assistance to manage the flow of information, students, and staff. Beyond the work of grading and returning graded work to students assisted by autograding software, there is a need to improve organization of students and teachers in office hours, and to distribute announcements, assignments, and learning materials to students as the number of students grows beyond traditional classroom capacities. As with autograding, student-facing infrastructure needs to scale as student demand can reach thousands of requests per hour, course rosters can grow beyond 1,000 students long, and there can be tens of thousands of forum posts per semester. These challenges, which can easily become immense administrative burdens at very large scale, have led to the development of specialized software to reduce the administrative overhead of running large courses. For instance, administering exams to over 1,600 students spread across as many as 20 rooms on campus has required the development of specialized software as well as more flexible, student-friendly policies and procedures to enable efficient support of hundreds of exam exceptions and accommodations each semester.

\section{Grading and Feedback}

Two of the most well-known grading and feedback web apps developed at UC Berkeley are Gradescope and OK. These two web apps stand out as particularly unique in how they have enabled new pedagogies and practices.

\subsection{Gradescope}

In Spring 2012, Pandagrader was conceived by the course staff teaching Introduction to Artificial Intelligence to streamline the process of grading paper exams. As instructors from other courses and institutions rapidly adopted the tool, in 2014, Pandagrader was incorporated as Gradescope.\footnote{\url{https://gradescope.com}}

The exam grading workflow begins with scanning and uploading exams to Gradescope. For very large courses with two, high-throughput copy machines, scanning all of the exams can take between 1--3 hours but yields significant time savings during the actual grading through Gradescope's fast online grading interface. Exam scores can then be turned around to students without returning physical papers: students view their graded exams online and see exactly which rubric items were applied. All grading and regrade requests are done over Gradescope's web interface that, by default, hides the identity of the student and the instructor from each other to minimize bias. The grading process is simpler for homework assignments as students submit assignments online themselves.

By moving the grading workflow online, course staff can increase their grading efficiency. Instead of shuffling papers, course staff assign rubric items to student submissions with only a single click or keystroke, and advance to the next submission with just another click or keystroke. Grading assignments according to an instructor-defined rubric improves transparency to students, helps ensure consistency between multiple graders, and makes it easy for the instructor to adjust point allocations or rubric item description post-hoc. Certain types of questions such as multiple choice or short answer blanks can be graded even more quickly with machine learning-assisted answer grouping where student submissions are automatically categorized into unique answer groups, allowing instructors to grade each group rather than each individual submission. Grading progress and statistics are computed on-the-fly, providing insight into aggregate and individual student success data. Gradescope's distributed grading system is also helpful for grading weekly problem sets. Students submit their work directly to Gradescope. Submissions can be graded online, enabling students to receive feedback the same week or even the day after submission depending on the length of the assignment and the amount of grading support.

The fast turnaround time also makes Gradescope a platform for providing formative feedback to students in two introductory CS courses at UC Berkeley. In these courses, students take a short paper quiz in their discussion or lab section during the day. After the section ends, the section TA scans their quizzes and uploads them to Gradescope. In the evening or later in the week, the course instructor and a handful of graders then grades all of the quizzes on Gradescope, identifying the entire class's common misconceptions, and returns personalized feedback to students via email. In end-of-semester course evaluations, students appreciated the additional feedback and found the system as a whole beneficial for their learning. Instructors and section TAs gain a detailed view of student performance previously unavailable in large courses while minimizing additional grading responsibilities. Using the misconceptions collected through weekly quizzes, instructors have released additional, targeted tutorials, practice materials, and study guides to help students improve before taking a high-stakes exams.

In 2017, Gradescope added support for autograding programming assignments. Gradescope's autograding platform is designed on top of container virtualization technology, allowing servers to start new autograder instances as needed and spread load across multiple machines. This ensures that students do not need to wait in a queue: new autograder instances can be provisioned as soon as students submit their assignments and, in times of greatest demand, servers can be dynamically added to the computing resource pool. This approach benefits instructors as they have full control over their choice of grading language and environment in the container. But it also presents a new cost as time needs to be invested in designing small programs or scripts that produce outputs following a specific autograder specification. Manual grading is also possible with grading rubrics and inline comments through a workflow similar to that of the homework and exam assignment types. The recent integration of a system for detecting software similarity has the potential to additionally simplify the workflow for identifying and understanding cases of over-collaboration.

\subsection{OK}

OK\footnote{\url{https://okpy.org}} is an online autograding platform developed by the course staff teaching introductory computer science at UC Berkeley in Fall 2014. Its primary users are introductory computer science and data science courses at UC Berkeley, each regularly serving enrollments 500--1,500 students per semester. The OK platform provides an alternative to Gradescope's autograding for programming assignments, though its approach favors much deeper integration with the course. In addition to a web interface for students to submit programming assignments, OK provides a Python client that boasts three key features over Gradescope's server-side grading.

\subsubsection{Student-Side Autograding}

With student-side autograding, students can run a suite of tests on their computer at any time, providing instantaneous code correctness feedback without needing to formally submit their assignment. These test suites are written in Python doctest format, a format students are familiar with from their practice using the interactive Python shell, which makes the tests and results easier for students to interpret than traditional unit tests. Complex integration tests can be written in separate files that are seamlessly integrated into the system \cite{Sridhara:2016}, appearing only after the submission passes targeted doctests. OK presents a simple interface to this feature so that students can autograde any part of the assignment without memorizing a complicated command.

Student-side autograding has resulted in a qualitative change in the way students approach problems as they now often use the autograder as part of a continuous edit-test feedback loop. For every change to a piece of code, students will re-run the autograder to see how the change affects the program. The immediate feedback then informs future planning, implementation, and debugging behaviors. This real-time feedback reduces frustration and builds student confidence by helping them make progress where they would normally get stuck. But there is an inherent risk to this approach. While students may be able to make more progress with the help of the student-side autograders, they may also grow more dependent on the feedback. Several introductory CS courses have experimented with velocity limiting, which limits autograder usage (either student-side or server-side) to a certain number of attempts that resets after a fixed amount of time. Students are encouraged to instead try out different debugging and problem-solving techniques such as developing their own examples and running through it on paper to check their understanding. This automated feedback has become an important cornerstone of introductory CS at UC Berkeley.

\subsubsection{Test Unlocking}

One of the risks of student-side autograding is that students can access tests without thinking through the problem, which can lead to a dependency on this development cycle. With test unlocking, the expected output of each doctest is stored as an encrypted string until it is successfully unlocked. Before students can run the student-side autograder, they must explore the problem by unlocking each automated test and determining the expected outputs by hand. Test unlocking reduced the number of conceptual questions (misunderstandings or clarifications of the problem), allowing instructors to spend more time assisting students with more involved questions: unlocking the tests helped students better understand the problem specifications and ``work through the thought process'' \cite{Basu:2015}. This result has been cited to provide metacognitive scaffolding \cite{Prather:2019}. Students later have the opportunity to write and run their own tests, reinforcing program understanding.

\subsubsection{Automatic Backups}

Each time the student-side autograder is invoked, student work is automatically backed up to the OK server along with metadata on their current progress. This submission and metadata collected by the automatic backup mechanism helps instructors answer questions about student learning. Together with the code backup, the OK client program records analytics such as the current question the student is working on, number of times the student-side autograder has been invoked thus far, and correctness. In addition, because the student-side autograder is run as part of a continuous edit-test feedback loop, the OK server aggregates a large number of intermediary student submissions. Instructors can then track student progress in aggregate and at the individual level to a high degree of detail. Frequent intermediary code snapshots have also benefited the research community as it has generated a massive dataset for education researchers to model student learning and performance \cite{Piech:2012, Wang:2017, Liao:2019}, deploy learning interventions at scale \cite{Sridhara:2016, Phothilimthana:2017}, understand excessive collaboration \cite{Yan:2018}, and propagate feedback at scale \cite{Yan:2019, Glassman:2015}. Students have the option to disable automatic backups and metadata collection but still run the student-side autograder by running the OK client with the appropriate command-line arguments.

\section{Managing Student Learning}

Beyond grading and feedback, many other components of a traditional lecture-format course such as office hours, exam scheduling, and discussion and lab scheduling are more difficult to organize at scale as they often require coordinating a large number of students.

\subsection{Office Hours}

Drop-in office hours in computer science courses have historically been organized as single-instructor events. The instructor announces their office hours time and place, and students visit office hours to interact with the instructor, often asking questions about assignments, concepts from lecture, as well as topics not directly related to the course such as undergraduate research opportunities and their own personal interests. However, as enrollments have increased, this model has proven increasingly difficult to scale especially as the number of homework questions has grown with the number of students. Questions left unanswered during discussion and lab section are clarified in office hours, leading to more demand for teaching assistance in office hours.

Until recently, this model was also true for teaching assistant office hours. In TA office hours, students mostly ask questions related to their assignments. TA office hours often used the same model with one TA walking around and answering each question. As more students arrive at office hours and the number of students waiting increases, students sign themselves up on a queue to reserve their place in line, ensuring that students who have waited the longest get helped next. While this model works up until about 10 students in office hours, as more students join the queue, there is pressure on the course staff to resolve all of the questions on the queue or risk ending office hours with several disappointed students. Overcrowded office hours are one of the highest-visibility issues for large lecture courses and a common source of student complaints.

Double, triple, and quadruple-staffing office hours has helped to alleviate supply and demand mismatches. Instead of office hours simply being the responsibility of a single TA, multiple TAs staff each office hour to meet student demand. However, one of the side effects of increasing the supply of TAs is that office hours can quickly grow out of control and unmanageable.  Students may jump ahead in the queue and get help from multiple different instructors, which not only takes teaching resources away from other students, but can also potentially shortcut the student's learning. Locating the next student in the queue is often a mess as instructors shout names across the room and need to navigate a maze of students to reach the next student on the queue.

Software can be used to streamline drop-in office hours by matching TAs to the next unresolved student help request through an online first-come, first-served queue. The Office Hours Queue web app, developed by undergraduate teaching assistants in the EECS Department on top of the OK platform, presents an interface for students and TAs to interact with the queue in much the same workflow as before. Students login and request help through the web app, specifying the assignment, question, location, and a brief description of the help request. TAs access the web app on their smartphones to view students on the queue. When a TA chooses to help the next student, the system marks the request as currently being helped, displays the name of the student to the TAs, and sends a notification to the student's device. The TA then works with the student to help resolve their question and make progress on their assignment. Finally, the TA marks the help request as resolved, or, if the help request was not resolved, returns the help request back to its original place on the queue. Similar systems have been deployed at peer institutions \cite{Smith:2017, MacWilliam:2013}. An introductory computer science student commented that,
\begin{quote}
    The additional office hour staff and expanded hours during [the final project] was great! I remember having to wait 2--3 hours and not being able to receive help for projects and assignments at the beginning of the school year because office hours were so packed and there wasn't really an orderly way to identify students who needed help.
\end{quote}
Other institutions have taken this model even further by moving office hours into highly-frequented, open spaces on campus for the convenience of students and course staff \cite{MacWilliam:2013}. In spite of these improvements, office hours remains oversubscribed, and finding spaces large enough and flexible enough for office hours remains a challenge.

This model for drop-in office hours treats student help requests as discrete, individual tickets. However, other models for office hours, such as The Tao of TALC, can also be highly productive. Instead of directly assisting students, The Tao of TALC encourages instructors to be guides wherein ``students drive as much as possible'' and the instructor acts ``more as a facilitator between the confused student and their `peer instructors'{}'' \cite{Astrachan:2007}. This has been used successfully both by instructors of small-group office hours and by TAs running large-group office hours. This group learning model has informed one of the features of the office hours queue, which enables a single instructor to help a group of students working on the same question all at once.

The use of online office hours queues has generated a large amount of data on office hours usage patterns \cite{Smith:2017}. Data dashboards for the office hours queue give instructors a view of which assignments students are asking questions about and when. Using this information, course staff have been able to shift resources and staff the most heavily-impacted office hours with more TAs where they were needed the most, further reducing student wait times. We have found that the majority of office hours are utilized by a very small minority of students: in a class of over 1,400 students, 50 students asked half of all the questions in office hours that semester.

\subsection{Online Course Delivery}

Course delivery is the process of offering a course to students. Office hours, lab, and discussion section are but a few ways students learn in large lecture courses. CS courses at UC Berkeley emphasize solving problems, which can occur through lecture, office hours, lab, and discussion section, but is often further emphasized in homework assignments and programming projects that combine multiple concepts presented in the course. As enrollments for several lower-division and upper-division CS courses regularly exceed the capacity of the largest lecture halls on campus, lectures in most CS courses are now webcasted. Making attendance at lecture, lab, and discussion section optional has had profound impacts on the way large lecture courses are run and has required special attention from the instructor.

In webcasted courses, it is much easier for students to fall behind and out-of-sync with content from lecture. Course policies need to be designed such that webcasted courses demand enough attention and consistent effort from students so that they stay on track. The assignment of frequent, small quizzes checks that students are keeping up with lecture. At UC Berkeley, simulcasts are not offered so live lecture recordings are often delayed by at least two hours. Nonetheless, even with interventions to keep students at pace with the course, it is often the case that many students will be at least a few days behind schedule at various times during the semester due to commitments from other courses taking priority.

In order to keep asynchronous classes up-to-speed, most large CS courses at UC Berkeley deliver their content via course websites or online discussion forums such as Piazza. The largest introductory CS courses update the front page of their website frequently with announcements so that students treat it as the definitive authority for the course, reducing student questions about assignments, deadlines, and exams. All materials in the course, including lecture videos, lecture notes, assignments, and readings, are always posted online, which enables students to choose how they would like to learn: either with a long-distance learning model, or by attending activities in a more traditional classroom setting.

Experienced TAs have also devised workarounds to make Piazza more robust for large courses.
\begin{quotation}
    Piazza will routinely be flooded with questions, especially near exams. Managing Piazza effectively does not necessarily mean keeping up with this flood of questions; it means directing the flow of questions efficiently so that students with similar questions can easily find previous students' questions and avoid asking repeat questions.

    Each week, keep a pinned Piazza thread for each separate homework question. This is the most important rule to follow, as homework questions will comprise the majority of all Piazza questions. Also, we recommend creating separate threads for each lecture note and discussion. Once these threads are in place, aggressively mark student questions as duplicates and move them to the appropriate threads so that all similar questions are in one place.

    To help students find these various threads, create a pinned master index that contains links to all of the above threads, as well as threads containing homework grade distributions, TA resources, weekly posts, events (guerrilla sections, exam review sessions, etc.), and other announcements. The index requires maintenance, so either consciously assign the management of the index as a TA duty, or make it understood within the team that each person who posts new material on the Piazza is also responsible for adding a link in the index.
\end{quotation}

Automation has also benefited online course delivery by streamlining administrative processes. In order to facilitate rapid updates, course websites are deployed using static website generators such as Jekyll for easy updating by course staff. Small programming questions and math problems are checked into a version-controlled question bank consisting of every question ever developed in the course so that developing a homework assignment can be as simple as specifying which questions to include and running a makefile to deploy the assignment to the website. TAs have even developed scripts to automatically extract screenshots and compose Piazza threads for each question. GitHub Classroom allows instructors to easily provision private GitHub repositories for student work and automatically configure instructor permissions. Some courses that require more advanced configuration have developed their own custom web scripts for provisioning private repositories. By automating these tasks, more time can be spent supporting student learning.

\subsection{Exam Administration}

Administering exams at scale is a particular pain point for large courses. One of the reasons for this is the number of students who need to be organized together, on-campus, at one time. In the largest introductory CS courses, it is common for exams to be simultaneously proctored across all of the largest lecture halls on campus, as well as spread across a dozen smaller classrooms. Traditionally, the course staff assigns students to exam rooms based on name or student identification number ranges. While, in theory, this system should allow the staff to distribute students however they like, large courses are hampered by the number of students requesting exam accommodations. In the largest introductory CS courses, there can be as many as 100 students requesting accommodations, which makes coordinating exam seating a challenge of its own.

Undergraduate teaching assistants in the EECS Department developed the Seating web app to solve this problem. The exam seating tool allows instructors to design seating charts, populate them with students based on their individual preferences, and send personalized emails to students with their particular exam room and exact seat location. Seating charts can be specified to assign students to every-other seat, or to skip entire rows of seats to make it easier for TAs to answer questions from students in the middle of a long row in a packed lecture hall.

Adopting this software has enabled large courses to provide better accommodations to all students, both students with and without documented disabilities. Students who are left-handed, or those who simply prefer to sit on a particular side of the room, near the aisle, or near the front can mark their preferences in a form. The Seating app will then take those preferences into account when randomly assigning a seat to the student. Since seats are assigned, it is easy to account for missing students and produces a paper trail making it more difficult for students to cheat on exams. Before, during, and after the exam, instructors can lookup any student on the seating chart and immediately identify adjacent students. When used with Gradescope, the Seating app makes it easier to compare suspicious student work against other students sitting in their vicinity.

Peer institutions have developed specialized computer-based testing facilities \cite{Nip:2018, West:2016} and restricted computing environments for test-taking \cite{Piech:2018} that have demonstrated benefits for student learning while reducing the amount of course staff resources spent administering exams.
\chapter{Support}
\label{chapter:support}

First popularized in the 1980s \cite{Reges:1988}, undergraduate students have been increasingly utilized in teaching positions at institutions of all sizes \cite{Roberts:1995, Reges:2003, Decker:2006, Dickson:2017, RetentionCS, Pon-Barry:2019}.
\begin{quote}
    Undergraduates have the potential to provide significant help and reduce the workload for graduate TAs and for faculty and other instructors. Undergraduates are typically paid per hour of effort, which can cost significantly less than graduate TAs. In addition, the shared experience of undergraduates may make them particularly attuned to understanding the problems and the challenges facing their peers around specific content. Furthermore, from a pedagogical perspective, peer teaching and evaluation can be valuable learning experiences in and of themselves, and can help empower students and build their confidence with the material. Finally, the undergraduate pool is larger than those of graduate students or faculty, and typically more diverse, presenting an opportunity for a more diverse set of instructors, which could contribute to a more inclusive culture. \cite{CSUndergraduateEnrollments}
\end{quote}

In the EECS Department at UC Berkeley, undergraduate teaching assistants (TAs) have been utilized since before 1987. One of the defining features of the undergraduate teaching program at UC Berkeley is the culture of student-directed innovation. Tools such as OK were developed by undergraduate teaching assistants to solve grading and feedback challenges. Situated within this existing infrastructure, the role of the instructor is part role model and part leader, with the goal of fostering a productive undergraduate teaching culture.

Students have multiple pathways to engage with the broader community beyond course staff. A constellation of extracurricular opportunities for students has developed over the past few years. Dedicated staff have created several services for the CS student community such as the CS Scholars program in which cohorts of 30 students take classes together for three semesters. With rising enrollments, more students than ever before are involved in EE and CS student organizations. The programs developed by staff and students for recruiting and retaining women in CS have been recognized by the National Center for Women \& Information Technology.\footnote{\url{https://www.ncwit.org/2019-ncwit-extension-services-transformation-next-award-recipients}}

\section{Undergraduate Teaching Assistants}

In the literature, it is common for the term Undergraduate Teaching Assistant to describe an undergraduate student teaching in a non-traditional role such as providing assistance to students in drop-in office hours or offering optional small-group or one-on-one tutoring. Notably, this definition refers to students who have a set of responsibilities that are distinct from graduate TAs. At UC Berkeley, undergraduate students hired as teaching assistants are responsible for the same set of duties as their graduate student counterparts: they teach weekly lab and discussion sections, grade assignments and exams, and assist with course delivery. These undergraduate students are officially hired under the title Graduate Student Instructor (GSI), though they are sometimes referred to as Undergraduate Student Instructors (UGSIs) to more accurately describe their academic standing.

EECS PhD students have a teaching requirement of 30 hours of service as a GSI, including 20 hours of service in an undergraduate course. However, the demand for TAs far exceeds the supply of graduate students. This is exacerbated by the fact that the vast majority of EECS PhD students are supported by research grants or fellowships that enable them to focus on their research. The hiring situation is particularly impacted in lower-division CS courses as graduate students often prefer to teach courses in their specific research area and taught by their faculty research advisors. As such, most introductory CS courses at UC Berkeley are taught with primarily undergraduate teaching assistants despite the fact that graduate TAs receive priority for these positions.

As the number of declared CS majors has increased, the demand for TAs in upper-division courses has also exploded beyond the supply of graduate TAs. Compared to peer institutions, the design of the undergraduate CS program makes it easier for the EECS Department to identify qualified undergraduates to staff upper-division courses and meet demand in spite of a shortage of interested graduate students. CS upper-division coursework is relatively flat with short prerequisite chains. Courses such as Introduction to Artificial Intelligence, Operating Systems, or Computer Security do not require any courses beyond the introductory course sequence. Students often satisfy the core introductory courses within two or three semesters so that, by the end of their second year, many students will have taken a couple upper-division CS courses that they can then teach over the remaining two years of their undergraduate degree program. Additionally, students are not required to complete a capstone project, leaving them more time to commit to coursework or extracurricular activities such as teaching or research.

In 2011, the largest introductory CS courses at UC Berkeley would hire 10 teaching assistants to serve classes with enrollments of about 350 students. Typical TA duties included both teaching and administrative responsibilities, usually split evenly across the entire course staff.
\begin{description}
\item[Teaching] Leading lab and discussion sections each week; holding weekly office hours; advising students; preparing for these teaching activities; and participating in weekly staff meeting.
\item[Administrative] Developing handouts, lab exercises, homeworks, and projects; grading assignments; handling accommodations for exceptional circumstances; managing announcements and student questions on the course forum; and proctoring and grading exams.
\end{description}

In recent years, however, grading and feedback tools such as Gradescope and OK have automated or streamlined many grading tasks. Tools have been developed to simplify traditionally expensive processes such as exam administration and assignment extensions. Furthermore, as course enrollments increase, the workload for certain aspects of course administration remain fixed. For example, the number of assignments and exams in the course is generally independent of the number of students enrolled in the course. In contrast, the teaching load grows linearly with respect to the number of students in the course.

Course staff composition has changed to reflect this new context. In a class of over 600 students with more than 20 TAs, there might be only a handful of 5--10 head TAs who are responsible for all of the course's administrative tasks. Instead of splitting tasks evenly, these 5--10 head TAs are each assigned one or two administrative responsibilities in addition to their regular teaching duties. There may be one or two TAs responsible for developing handouts, lab exercises, homeworks, and projects, which allows them to become domain experts in developing assignments for the course and maintaining a high quality of assignments with fewer bugs and greater consistency. This shift allows the remaining TAs to focus on teaching their students as effectively as possible. The number of these teaching-focused TA positions can be scaled at the same rate as course enrollment without significantly affecting course administration activities. Managing this greater number of teaching-focused TAs has become an administrative responsibility in and of itself so there may also be a head TA whose duty is to manage the course staff, communicate expectations, announce upcoming activities, and improve the quality of teaching.

\section{Center for Student Affairs}

The EECS Department served over 27,000 student enrollments across all course offerings during the 2018 academic year. Managing this number of students presents an administrative challenge for operating the program at scale, and can easily create feelings of anonymity among students, harming recruitment and retention efforts. The Center for Student Affairs (CSA), an EECS staff unit that provides several functions for undergraduate and graduate CS education, has developed a number of programs and solutions to tackle these challenges.

Starting Fall 2013, the CS Scholars Program,\footnote{\url{https://eecs.berkeley.edu/cs-scholars}} based on student retention theories of first year college students \cite{Tinto:1987, Terenzini:1980} and minority engineering students \cite{Treisman:1992}, is one such solution.
\begin{quote}
    CS Scholars is a first-year student support program intended to serve those from under-represented communities who have had little or no exposure to Computer Science. A learning community, CS Scholars integrates several components of support to meet the academic, social, and developmental needs of students intending to study Computer Science. Those components include:
    \begin{itemize}
    \item Cohort-style course discussions
    \item CS Scholars only seminars for personal and professional development
    \item Solidarity and community building activities
    \item Dedicated CS Scholars Advising
    \end{itemize}
\end{quote}
Data analysis has shown that the CS Scholars cohorts outperform students in the general population by 10--20\%, and students maintain a higher GPA than the overall class. In earlier cohorts, among students who identify as having no prior programming experience, CS Scholars had a 0.3 GPA advantage over non-scholars, and a greater difference for students who self-identified as female. The CSA-led EECS Resiliency Project is another retention initiative, which draws attention to stories from students and faculty who struggled with computer science at some point in their lives but persevered through those experiences of failure.

To diversify participation in and access to research experiences and graduate work in computer science, the CSA developed the Summer Undergraduate Program in Engineering Research at Berkeley (SUPERB),\footnote{\url{https://eecs.berkeley.edu/resources/undergrads/research/superb}} an NSF-funded Research Experience for Undergraduates (REU) program. Participants include junior and senior undergraduate students at Berkeley or elsewhere, and each participant receives faculty mentorship, graduate student support, and graduate school advising. 95\% of the students who participated in SUPERB continued to graduate school in STEM fields \cite{Alivisatos:2017}.

Due to the large number of students in the EECS major or considering declaring the LSCS major, most undergraduate advising is provided by professional EECS and LSCS major advisors. The advising staff assists with student questions and concerns including those related to the CS degree programs, coursework, undergraduate research, as well as students' broader plans and how they might fit into their life or career goals. The advising staff also manages a team of undergraduate peer advisors.

Getting into CS courses has become an often-cited grievance for undergraduates enrolled in universities, both large and small, across the nation. One of the CSA's functions is to coordinate between faculty and students to ensure that teaching supply is properly calibrated to meet enrollment demands, so the CSA has dedicated staff members for managing course scheduling and enrollment. Course capacity in the EECS Department is primarily limited by availability of classrooms and teaching assistants. Allocation of most discussion classrooms and large lecture halls involves close cooperation with central campus administrative staff. However, campus spaces do not provide enough capacity for all CS courses, so space often needs to be found within EECS-managed buildings. This is complicated by the fact that many spaces are already earmarked for strictly research purposes or strictly academic purposes. The challenge of efficiently allocating the remaining shared spaces is further exacerbated by enrollment growth as research functions compete with rapidly-growing academic functions. Increasing enrollments has also increased the number of teaching assistant hires, which has resulted in a significantly enlarged payroll that, unfortunately, does not increase at a rate sufficient to meet teaching needs, let alone match enrollment trends. Additionally, hiring more TAs requires additional coordination with campus training for first-time GSIs, as well as department-level and course-level preparation (\autoref{chapter:preparation}).

\section{Near-Peer Student Mentors}

As of 2018, there are 42 student organizations officially registered with the EECS Department, many of which host events and provide services to the broader CS community, such as:
\begin{description}
\item[Mentorship] Student organizations provide mentorship opportunities by hosting one-on-one or small-group mentoring sessions, blending academic support with a sense of community.
\item[Invited Speakers] External speakers and alumni give talks on topics including diversity in tech, overcoming adversity, and well-being, as well as workshops on bias, equity, and inclusion.
\item[Industry Events] With a student organization as their sponsor, employers can host info sessions, tech talks, or other events such as puzzle hunts or trivia nights to network with students.
\end{description}
One of the unusual features of the EECS Department is the amount and diversity of student-driven, near-peer mentorship opportunities available to students. In the near-peer mentor model, mentors are only a couple years more senior than their mentees. Near-peer mentoring ``provides younger students with a positive, inspiring experience to learning about computing from college near-peer mentors,'' and ``helps students feel like they belong in CS, especially if their mentors have backgrounds or experiences similar to their own'' \cite{RetentionCS}.

One such mentorship program is CS Kickstart.\footnote{\url{https://cs-kickstart.berkeley.edu}}
\begin{quote}
   CS Kickstart is a week-long program open to any incoming UC Berkeley students that introduces them computer science while meeting other computer science students and professionals. This program primarily targets women who are interested in the fields of science, technology, engineering, and math. Participants get hands-on experience in programming introducing them to the creativity and diversity of computer science. Participants also get the opportunity to visit tech companies in the Bay Area to see what life is like for computer scientists in industry. For several years it served 25 incoming students, but recently this doubled. It draws almost all of its support from industry and individual donors. \cite{Alivisatos:2017}
\end{quote}
The 2019 cohort will consist of about 50 participants. The program is organized by a group of undergraduate and graduate women in computer science, and is offered free to participants despite housing, transportation, and activity costs thanks to industry sponsors. As a result of the program, 96 percent of participants felt more prepared to take their first CS course at Berkeley, and 95 percent had a greater motivation to pursue computer science.

Once students are on campus, there are several student organizations forming communities around various identities or affinity groups, many of which offer mentorship programs of their own. Serving the woman-identifying EECS community is the Association of Women in Electrical Engineering and Computer Science (AWE).
\begin{quote}
    The AWE Mentorship Program provides a framework for EE and CS women to develop and sustain mentoring relationships by matching incoming students with upper division women. As new students, mentees connect with their mentors at the beginning of the school year, receiving personalized academic and social help when needed. Throughout the academic year, mentees receive advice, encouragement, information, and insight from experienced peers. Mentors, in turn, gain satisfaction and knowledge from guiding fellow students while fostering a sense of community.\footnote{\url{https://eecs.berkeley.edu/resources/undergrads/eecs/women/mentoring}}
\end{quote}
Similarly, the Society of Women Engineers provides mentorship to the broader community of women in all kinds of Engineering, and the more recent FEMTech student organization engages with the broader campus community by providing outreach and mentorship activities such as FEMTech Launch, which ``provides office hours, extra help, and weekly tutoring sessions specifically geared towards women and underrepresented minorities in lower level CS courses.''\footnote{\url{https://femtechberkeley.com/index.php/education/}} Honor societies such as Eta Kappa Nu (HKN) and Upsilon Pi Epsilon (UPE) offer free drop-in tutoring to the EECS undergraduate community across a majority of the undergraduate coursework.

Computer Science Mentors (CSM)\footnote{\url{https://csmentors.berkeley.edu}} is a student organization that, like other programs, offers academic support together with the community-building benefits of near-peer mentorship, but is offered at large scale, serving nearly 2,000 students per semester across 6 introductory computer science and electrical engineering courses. A typical mentoring group consists of 4--6 students and 1 near-peer mentor. The mentor facilitates student discussions and group work with a focus on mastery learning. Mentors adapt each session to meet the group's needs, drawing on additional examples to clarify concepts and build student confidence. Over the course of the semester, the mentor gets to know each student on an individual basis, and students grow more comfortable with each other too. The development of these relationships makes it easier for the mentor to keep in touch with their students by setting up individual check-ins in addition to the group sessions, sharing their experiences and study advice, and referring students to free tutoring services offered by other members of the EECS community. Participation in CSM small-group mentoring has been shown to have a significant positive association with exam scores. Organizing, preparing, and mentoring the mentors has become a challenge of its own (\autoref{chapter:preparation}).
\chapter{Preparation}
\label{chapter:preparation}

Utilizing undergraduates in teaching positions is not without its risks.
\begin{quote}
    Undergraduates who are unclear on the material may cause confusion among their peers. In addition, not all undergraduates have the knowledge or maturity to successfully teach, assess, or mentor their peers, or understand conflict-of-interest situations. If poorly implemented or not properly supervised, this approach can place additional strain on course instructors. \cite{CSUndergraduateEnrollments}
\end{quote}
Preparation is especially important as the program expands in size and hires more undergraduate TAs to support large enrollment courses in both lower-division and upper-division courses.

\section{Introduction to Teaching Computer Science}
\label{section:cs370}

\begin{quotation}
    [CS 370: Introduction to Teaching Computer Science] is a course designed help aspiring teachers hone their teaching skills, become a part of the teaching community here at UC Berkeley, and expose them to the foundations of computer science pedagogy. Students in this class will receive first-hand experience through one-on-one tutoring and an enriched teaching knowledge through research-based pedagogical studies.

    CS 370 has three key components that distinguish it from other pedagogical courses. First, we cover student interactivity and teaching in one-on-one settings. This is applicable to all levels of teachers [\dots\unkern] since one-on-one interactions are a critical component of all teaching experiences. Next, we cover group teaching through in-class demonstrations, as mastering pacing and understanding the individualities of students in a group setting is key to being a successful TA. Last, we socratically discuss current issues in CS pedagogy, including atmosphere-related questions such as: underrepresentation, stigmas associated with computer science, the issue of prior experience, and how these factors heavily influence student learning.\footnote{\url{http://inst.eecs.berkeley.edu/~cs370/policies.html}}
\end{quotation}

Students are introduced to pedagogical concepts during an 80-minute seminar each week that includes discussion of ideas and reflection on their teaching experiences in small groups. Outside of the classroom, students read scholarly articles on the practice and theory of teaching computer science, host three, hour-long one-on-one tutoring sessions per week, and reflect on their tutoring as part of a weekly written assignment. Combining theory and practice together helps students learn and retain material, treating topics taught in class as a frame for questions brought up in self-reflections on their teaching experiences. Group discussions are facilitated by experienced TAs whose experience students identify with and more closely relate. To facilitate discussion, these weekly seminars are held in an active learning classroom with students seated around tables and facing each other rather than the front of a lecture hall. Between the weekly seminar, tutoring, tutoring preparation, tutoring reflection, and weekly assignments, CS 370 is a total commitment of 9 hours per week.

CS 370 was designed originally as a course to prepare and engage new teachers, which influenced its decision to use one-on-one tutoring as the context for teaching practicum. Unlike other programs at peer institutions, a large number of aspiring undergraduate student-teachers take the course before they become TAs. This results in a diversity of students composed of first-year students who just recently took the courses they want to someday teach as well as older, second or third-year students, which makes for engaging conversations as their different experience levels provide greater opportunities to learn from each other. More-experienced student-teachers in the group and the experienced TA facilitators can chime in and provide nuanced viewpoints to questions less-experienced teachers might have about teaching one-on-one or leading small groups.

This design of CS 370 has a number of consequences that has made it particularly well-suited for preparing undergraduate TAs. First, the outline of topics includes concerns that are especially important for teaching at the undergraduate level such as diversity, unconscious bias, and tackling misconceptions. CS 370 is complemented by CS 375, which is geared toward a graduate student audience and, notably, includes coverage of topics such as developing course syllabi, exam problems or rubrics, and student surveys, all of which are tasks that concern head TAs responsible for the administrative component of a course but not necessarily teaching-focused TAs. CS 370, CS 375, as well as other pedagogy courses at peer institutions have also found success running the course in a workshop style with a significant portion of the materials presented at the beginning of the semester to maximize their effect on teaching, and then later fading away to more infrequent check-ins later in the semester \cite{Roberts:1995}.

\section{Mentoring at Scale}

Near-peer student mentors, such as the students who lead small-group sessions for CSM, are organized into a family system to prepare for their weekly group sessions and build community. Like TA families, mentors are grouped into families of about 6 mentors, each consisting of two experienced senior mentors and about four less-experienced junior mentors. In addition to providing feedback, checking-in, and bonding over social events, mentors also meet together regularly for one hour each week to prepare for the upcoming week's group sessions with mentees. Family meetings are a mix between active problem-solving, 3-minute teaching demonstrations, real-time critiques, and moments of written self-reflection. In these weekly family meetings, senior mentors lead and facilitate group discussions with junior mentors about the challenges and pitfalls of upcoming concepts, and assist mentors in personalizing their session to meet their mentees' needs. In order to make this hour effective, junior mentors prepare for the family meetings in advance by spending half an hour reviewing concepts in advance and preparing a mental outline of the lesson they have in mind for their session.

Most mentors only lead one or two group mentoring sessions. Since these mentoring sections only consist of 4--6 students each, for larger classes, over 100 sections are offered each week. In order to support this structure, CSM delegates the task of organizing mentors to the course coordinators, highly-experienced mentors who manage the entire operation. Course coordinators play a similar role as Section Leader Coordinators and Meta-TAs implemented at peer institutions \cite{Reges:2003, Roberts:1995, Reges:1988}. They hold weekly meetings with all of the senior mentors to prepare content for the family meetings and group mentoring sessions and assist the senior mentors in preparing for facilitating their own family meetings.

\section{Course-Specific Preparation}

Large classes make it harder for instructors to provide individual feedback to students. Likewise, large course staffs make it harder for instructors to provide personalized mentorship to their TAs. As course staffs have grown beyond 20, 30, 40, and even 50 TAs, several CS courses have begun grouping their course staff members into smaller families as well. Each TA family consists of 4--6 TAs with a mix of experienced and inexperienced teachers. As part of their preparation duties, TAs are occasionally expected to shadow and provide feedback to other family members to improve their teaching. Mirroring mentoring families, lead TAs check-in with their family members throughout the semester and organize occasional social outings with the entire group, building a community between undergraduate TAs.

In addition to the formal CS 370 pedagogy course and the more informal family system, undergraduate student-teachers also receive support and mentorship at the course level. The instructor of record and more-experienced TAs will often share their preparation materials, refine assignment guides, discussion walkthroughs, and other documents designed to support newer teachers. Discussions handouts are often reused between semesters so course staff share potential ways of teaching the concepts. Assignment guides provide answers to frequently-asked questions, identify common student bugs and their fixes, and suggest relevant connections to previous concepts and prerequisites to bridge knowledge gaps.

It is also common for course staff to run their own preparation sessions at the beginning of each semester to on-board new course staff members, set expectations, and provide course-specific guidance. Topics include preparing for discussion, lab, and office hours; modeling behavior and setting student expectations for the course's pace, format, and recommended learning strategy; course-specific resources and policies that need to be shared with students early in the semester; and upcoming changes for the current offering of the course. The course staff set four ground rules for one-on-one interactions in office hours:
\begin{enumerate}
    \item If you don't know what to do, ask.
    \item Be sensitive because learning computer science can be hard.
    \item Let the student drive.
    \item Do not give away the answer, if you can help it.
\end{enumerate}
These conversations are continued throughout the semester during weekly staff meetings where the entire course staff meets to make decisions on open administrative questions, give and receive feedback on new ideas or proposals, and plot out the next few weeks' content in the course. At the final meeting, time is set aside for course staff to reflect on the entire semester as a whole and determine where improvements can be made to assignments, teaching, policies, and the overall design of the course.

In addition, experienced TAs propose an idealized assignment help workflow to normalize lab and office hours expectations across the course staff. The goal of this workflow is to reduce the risk of providing too much assistance, which can harm students as they grow dependent on the guidance and are unable to solve problems on their own \cite{Smith:2017}. For programming courses, this workflow starts with understanding the question since students often miss important details when focused on the problem. The TA is directed to sit down beside the student, ensuring that their eye-lines match, and introduce themselves and learn the student's name. These practices help to build trust and rapport between the student and the TA, particularly if the student and the TA meet again in lab or office hours. The next step is to ask the student to describe their problem in their own words. This gives the TA time to skim the student's code and verify that the student's explanation matches their code, and later work with the student to identify the source and cause of the problem. After the student gains an understanding of the problem, the TA works with the student to formulate a plan to resolve the problem, and then gives the student time to solve it on their own. After about 10 minutes, during which the TA helps another student, the TA returns to check back in on the student's progress. These last few practices give the student space to work on the problem on their own and encourages them to build independence. Rather than sitting with the student and solving their problems for them, the TA's goal is to have the student in a better position to solve the problem independently.

\chapter{Discussion}
\label{chapter:discussion}

While these methods have enabled CS courses at UC Berkeley to scale to meet both CS major and non-major student demand, the system of incentives, particularly the LSCS 3.3 GPA cap, strains relationships between instructors and students. Implementing student-friendly course policies, designing collaborative assignments, and encouraging students to take advantage of mentorship opportunities can significantly improve the student learning experience but is ultimately contrary to the message sent by the GPA cap. There is little room for failure: students who struggle in one or two of the introductory courses face an uphill battle to make it to the GPA cap. This is in spite of the limited evidence that, when given a second chance, students are able to make a remarkable improvement. In Spring 2016, students who were given the option to receive a failing grade in introductory Data Structures and retake the course in a later semester made an average improvement of +2.54 grade bins over the grade they would have received in Spring 2016.

The EECS Department has made significant gains in improving the gender diversity of its undergraduate student population, receiving recognition by the National Center for Women \& Information Technology (NCWIT)\footnote{\url{https://www.ncwit.org/2019-ncwit-extension-services-transformation-next-award-recipients}} as well as local news media\footnote{\url{https://www.mercurynews.com/2018/04/16/forget-techs-bad-bros-stanford-berkeley-boost-female-computing-grads/}} for its achievements. However, there is still much work to be done to encourage participation from a broader population of students. For some students, the time, energy, and stress necessary to meet the GPA cap makes the major unattractive. Other students may not have the confidence to pursue the major despite interest and academic preparation. In order to grow capacity while maintaining an inclusive student culture, it is important to take into account the entire system of incentives and punishments. Policies such as the GPA cap have ripple effects as student perceptions of the program on campus are shaped by its reputation of being highly rigorous, demanding, and stressful. Students may not feel comfortable if they see the program---including its faculty, staff, and students---as competitive in spite of their best efforts to design collaborative and supportive learning experiences. When a potential student's sensibilities do not line up with these perceived values, students may feel excluded from CS even if they could otherwise be successful computer scientists.

Even with technology and a large number of highly motivated support staff, faculty teaching load remains a significant burden, particularly for new tenure-track assistant professors who also need to balance their research output and tenure priorities with teaching. In particular, the faculty find larger enrollments have resulted in greater administrative workload, one that has not yet been fully displaced by head TAs despite all of the preparation and software in place to support them. Furthermore, courses that rely on individually-personalized projects or deep-feedback assignments do not easily fit into this framework of automation solutions. As a consequence, many faculty now choose to co-teach courses, which reduces workload but restricts course offerings.

Additionally, the use of some automation has the potential to impact student behavior in unexpected ways. The office hours queue, for example, compartmentalizes assignment help and student questions into individual tickets that are then resolved one-by-one. This model works in introductory courses due to the large number of TA office hours, and because assignments are typically scaffolded to help students make progress and facilitate efficient resolution in office hours. But the kinds of debugging and self-regulation practices acquired through these introductory CS office hours don't necessarily prepare students for upper-division coursework where the questions are more open-ended and the debugging processes much less clear.

Designing a CS program that scales requires cooperation from all levels of campus, including the students, staff, faculty, the department, the college, and the administration. While this report focuses on recent developments, a culture of innovation by undergraduate TAs has long existed in the EECS Department since their introduction in the 1980s, and by graduate students even before then. Each class of students is supported by the preceding class of students who teach section each week and serve as role models. Staff have worked closely with students and faculty to develop novel solutions to challenges of teaching CS at scale while advising triple the number of students from just a decade ago. Faculty have made sacrifices to teach at this scale, often teaching courses double their expected teaching load due to exploding enrollments. The College of Engineering, Graduate Division, and campus administration have supported the program by committing additional faculty slots, expanding advising support, and expanding course enrollments by funding additional TAs.

However, the program still faces a number of budgetary shortfalls as campus Temporary Academic Support (instructional support) has not kept up with the unprecedented growth of the program. Contributions from private and industry donors have enabled the department to continue funding more TAs, opening more sections, and reaching more students in spite of structural deficits and budget cuts at the university and state level. This external investment has fueled the development of automation solutions, support initiatives, and preparation processes that have made the UC Berkeley EECS Department a national model for teaching CS at scale.

\bookmarksetup{startatroot}
\printbibliography[heading=bibintoc]

@online{Roberts:2018,
 author = {Roberts, Eric},
 title = {Resources for the CS Capacity Crisis},
 year = {2018},
 urldate = {2019-04-01},
 url = {https://cs.stanford.edu/people/eroberts/ResourcesForTheCSCapacityCrisis/},
}

@online{Roberts:2016,
 author = {Roberts, Eric},
 title = {A History of Capacity Challenges in Computer Science},
 year = {2016},
 urldate = {2019-04-01},
 url = {https://cs.stanford.edu/people/eroberts/CSCapacity/},
}

@report{Curtis:1982,
 author = {Curtis, Kent K.},
 title = {Computer Manpower --- Is There a Crisis?},
 institution = {National Science Foundation},
 address = {Washington, D.C., USA},
 year = {1982},
 url = {https://cs.stanford.edu/people/eroberts/Curtis-ComputerManpower/},
}

@report{GenerationCS,
 author = {{Computing Research Association}},
 title = {Generation CS: Computer Science Undergraduate Enrollments Surge Since 2006},
 year = {2017},
 url = {https://cra.org/data/Generation-CS/},
}

@report{TaulbeeSurvey2018,
 author = {Zweben, Stuart and Bizot, Betsy},
 title = {2018 CRA Taulbee Survey},
 year = {2019},
 url = {https://cra.org/resources/taulbee-survey/},
}

@report{CSforAll2018,
 author = {Delyser, Leigh A. and Goode, Joanna and Guzdial, Mark and Kafai, Yasmin and Yadav, Aman},
 title = {Priming the Computer Science Teacher Pump: Integrating Computer Science Education into Schools of Education},
 year = {2018},
 url = {http://www.computingteacher.org/2018}
}

@report{StateofCS2018,
 title = {2018 State of Computer Science Education},
 year = {2018},
 url = {https://advocacy.code.org/},
}

@techreport{Swamy:2018,
 author = {Swamy, Vinitra},
 title = {Pedagogy, Infrastructure, and Analytics for Data Science Education at Scale},
 institution = {EECS Department, University of California, Berkeley},
month = may,
 year = {2018},
 url = {https://www2.eecs.berkeley.edu/Pubs/TechRpts/2018/EECS-2018-81.html},
 number = {UCB/EECS-2018-81},
}

@book{CSUndergraduateEnrollments,
 author = {{National Academies of Sciences, Engineering, and Medicine}},
 title = {Assessing and Responding to the Growth of Computer Science Undergraduate Enrollments},
 isbn = {978-0-309-46702-5},
 doi = {10.17226/24926},
 url = {https://www.nap.edu/catalog/24926/assessing-and-responding-to-the-growth-of-computer-science-undergraduate-enrollments},
 year = {2018},
 publisher = {The National Academies Press},
 address = {Washington, DC}
}

@report{DiversityGapsCS,
 author = {{Google Inc. and Gallup Inc.}},
 title = {Diversity Gaps in Computer Science: Exploring the Underrepresentation of Girls, Blacks and Hispanics},
 year = {2016},
 url = {http://goo.gl/PG34aH},
}

@book{Tinto:1987,
 title = {Leaving college: Rethinking the causes and cures of student attrition},
 author = {Tinto, Vincent},
 year = {1987},
 publisher = {ERIC},
}

@unpublished{Alivisatos:2017,
 title = {STEM and Computer Science Education: Preparing the 21st Century Workforce},
 author = {Paul Alivisatos},
 year = {2017},
 url = {https://docs.house.gov/meetings/SY/SY15/20170726/106330/HHRG-115-SY15-Wstate-AlivisatosA-20170726.pdf},
 note = {Research and Technology Subcommittee, House Committee on Science, Space, and Technology},
}

@article{Terenzini:1980,
 ISSN = {03610365, 1573188X},
 URL = {http://www.jstor.org/stable/40195370},
 author = {Patrick T. Terenzini and Ernest T. Pascarella},
 journal = {Research in Higher Education},
 number = {3},
 pages = {271--282},
 publisher = {Springer},
 title = {Toward the Validation of Tinto's Model of College Student Attrition: A Review of Recent Studies},
 volume = {12},
 year = {1980},
}

@article{Treisman:1992,
 ISSN = {07468342, 19311346},
 URL = {http://www.jstor.org/stable/2686410},
 author = {Uri Treisman},
 journal = {The College Mathematics Journal},
 number = {5},
 pages = {362--372},
 publisher = {Mathematical Association of America},
 title = {Studying Students Studying Calculus: A Look at the Lives of Minority Mathematics Students in College},
 volume = {23},
 year = {1992},
}

@article{Cohoon:2002,
 author = {Cohoon, J. McGrath},
 title = {Recruiting and Retaining Women in Undergraduate Computing Majors},
 journal = {SIGCSE Bull.},
 issue_date = {June 2002},
 volume = {34},
 number = {2},
 month = jun,
 year = {2002},
 issn = {0097-8418},
 pages = {48--52},
 numpages = {5},
 url = {http://doi.acm.org/10.1145/543812.543829},
 doi = {10.1145/543812.543829},
 acmid = {543829},
 publisher = {ACM},
 address = {New York, NY, USA},
}

@inproceedings{Babes-Vroman:2017,
 author = {Babes-Vroman, Monica and Juniewicz, Isabel and Lucarelli, Bruno and Fox, Nicole and Nguyen, Thu and Tjang, Andrew and Haldeman, Georgiana and Mehta, Ashni and Chokshi, Risham},
 title = {Exploring Gender Diversity in CS at a Large Public R1 Research University},
 booktitle = {Proceedings of the 2017 ACM SIGCSE Technical Symposium on Computer Science Education},
 series = {SIGCSE '17},
 year = {2017},
 isbn = {978-1-4503-4698-6},
 pages = {51--56},
 numpages = {6},
 url = {http://doi.acm.org/10.1145/3017680.3017773},
 doi = {10.1145/3017680.3017773},
 acmid = {3017773},
 publisher = {ACM},
 address = {New York, NY, USA},
}

@inproceedings{Newhall:2014,
 author = {Newhall, Tia and Meeden, Lisa and Danner, Andrew and Soni, Ameet and Ruiz, Frances and Wicentowski, Richard},
 title = {A Support Program for Introductory CS Courses That Improves Student Performance and Retains Students from Underrepresented Groups},
 booktitle = {Proceedings of the 45th ACM Technical Symposium on Computer Science Education},
 series = {SIGCSE '14},
 year = {2014},
 isbn = {978-1-4503-2605-6},
 pages = {433--438},
 numpages = {6},
 url = {http://doi.acm.org/10.1145/2538862.2538923},
 doi = {10.1145/2538862.2538923},
 acmid = {2538923},
 publisher = {ACM},
 address = {New York, NY, USA},
}

@article{Narayanan:2018,
 author = {Narayanan, Sathya and Cunningham, Kathryn and Arteaga, Sonia and Welch, William J. and Maxwell, Leslie and Chawinga, Zechariah and Su, Bude},
 title = {Upward Mobility for Underrepresented Students: A Model for a Cohort-based Bachelor's Degree in Computer Science},
 journal = {ACM Inroads},
 issue_date = {June 2018},
 volume = {9},
 number = {2},
 month = apr,
 year = {2018},
 issn = {2153-2184},
 pages = {72--78},
 numpages = {7},
 url = {http://doi.acm.org/10.1145/3210555},
 doi = {10.1145/3210555},
 acmid = {3210555},
 publisher = {ACM},
 address = {New York, NY, USA},
}

@inproceedings{Rheingans:2018,
 author = {Rheingans, Penny and D'Eramo, Erica and Diaz-Espinoza, Crystal and Ireland, Danyelle},
 title = {A Model for Increasing Gender Diversity in Technology},
 booktitle = {Proceedings of the 49th ACM Technical Symposium on Computer Science Education},
 series = {SIGCSE '18},
 year = {2018},
 isbn = {978-1-4503-5103-4},
 pages = {459--464},
 numpages = {6},
 url = {http://doi.acm.org/10.1145/3159450.3159533},
 doi = {10.1145/3159450.3159533},
 acmid = {3159533},
 publisher = {ACM},
 address = {New York, NY, USA},
}

@inproceedings{Maher:2015,
 author = {Maher, Mary Lou and Latulipe, Celine and Lipford, Heather and Rorrer, Audrey},
 title = {Flipped Classroom Strategies for CS Education},
 booktitle = {Proceedings of the 46th ACM Technical Symposium on Computer Science Education},
 series = {SIGCSE '15},
 year = {2015},
 isbn = {978-1-4503-2966-8},
 pages = {218--223},
 numpages = {6},
 url = {http://doi.acm.org/10.1145/2676723.2677252},
 doi = {10.1145/2676723.2677252},
 acmid = {2677252},
 publisher = {ACM},
 address = {New York, NY, USA},
}

@inproceedings{Malan:2010,
 author = {Malan, David J.},
 title = {Reinventing CS50},
 booktitle = {Proceedings of the 41st ACM Technical Symposium on Computer Science Education},
 series = {SIGCSE '10},
 year = {2010},
 isbn = {978-1-4503-0006-3},
 pages = {152--156},
 numpages = {5},
 url = {http://doi.acm.org/10.1145/1734263.1734316},
 doi = {10.1145/1734263.1734316},
 acmid = {1734316},
 publisher = {ACM},
 address = {New York, NY, USA},
}

@inproceedings{Porter:2013,
 author = {Porter, Leo and Bailey Lee, Cynthia and Simon, Beth},
 title = {Halving Fail Rates Using Peer Instruction: A Study of Four Computer Science Courses},
 booktitle = {Proceeding of the 44th ACM Technical Symposium on Computer Science Education},
 series = {SIGCSE '13},
 year = {2013},
 isbn = {978-1-4503-1868-6},
 pages = {177--182},
 numpages = {6},
 url = {http://doi.acm.org/10.1145/2445196.2445250},
 doi = {10.1145/2445196.2445250},
 acmid = {2445250},
 publisher = {ACM},
 address = {New York, NY, USA},
}

@inproceedings{Guo:2013,
 author = {Guo, Philip J.},
 title = {Online Python Tutor: Embeddable Web-based Program Visualization for CS Education},
 booktitle = {Proceeding of the 44th ACM Technical Symposium on Computer Science Education},
 series = {SIGCSE '13},
 year = {2013},
 isbn = {978-1-4503-1868-6},
 pages = {579--584},
 numpages = {6},
 url = {http://doi.acm.org/10.1145/2445196.2445368},
 doi = {10.1145/2445196.2445368},
 acmid = {2445368},
 publisher = {ACM},
 address = {New York, NY, USA},
}

@inproceedings{Hug:2015,
 author = {Hug, Josh and Garcia, Daniel D.},
 title = {Handling Very Large Lecture Courses: Keeping the Wheels on the Bus (Abstract Only)},
 booktitle = {Proceedings of the 46th ACM Technical Symposium on Computer Science Education},
 series = {SIGCSE '15},
 year = {2015},
 isbn = {978-1-4503-2966-8},
 pages = {697--697},
 numpages = {1},
 url = {http://doi.acm.org/10.1145/2676723.2691867},
 doi = {10.1145/2676723.2691867},
 acmid = {2691867},
 publisher = {ACM},
 address = {New York, NY, USA},
}

@inproceedings{Hug:2017,
 author = {Hug, Josh and Lee, Cynthia},
 title = {Handling Very Large Lecture Courses: Keeping the Wheels on the Bus III (Abstract Only)},
 booktitle = {Proceedings of the 2017 ACM SIGCSE Technical Symposium on Computer Science Education},
 series = {SIGCSE '17},
 year = {2017},
 isbn = {978-1-4503-4698-6},
 pages = {725--725},
 numpages = {1},
 url = {http://doi.acm.org/10.1145/3017680.3022374},
 doi = {10.1145/3017680.3022374},
 acmid = {3022374},
 publisher = {ACM},
 address = {New York, NY, USA},
}

@inproceedings{Alvarado:2017,
 author = {Alvarado, Christine and Minnes, Mia and Porter, Leo},
 title = {Micro-Classes: A Structure for Improving Student Experience in Large Classes},
 booktitle = {Proceedings of the 2017 ACM SIGCSE Technical Symposium on Computer Science Education},
 series = {SIGCSE '17},
 year = {2017},
 isbn = {978-1-4503-4698-6},
 pages = {21--26},
 numpages = {6},
 url = {http://doi.acm.org/10.1145/3017680.3017727},
 doi = {10.1145/3017680.3017727},
 acmid = {3017727},
 publisher = {ACM},
 address = {New York, NY, USA},
}

@inproceedings{Minnes:2018,
 author = {Minnes, Mia and Alvarado, Christine and Porter, Leo},
 title = {Lightweight Techniques to Support Students in Large Classes},
 booktitle = {Proceedings of the 49th ACM Technical Symposium on Computer Science Education},
 series = {SIGCSE '18},
 year = {2018},
 isbn = {978-1-4503-5103-4},
 pages = {122--127},
 numpages = {6},
 url = {http://doi.acm.org/10.1145/3159450.3159601},
 doi = {10.1145/3159450.3159601},
 acmid = {3159601},
 publisher = {ACM},
 address = {New York, NY, USA},
}

@inproceedings{Kay:1998,
 author = {Kay, David G.},
 title = {Large Introductory Computer Science Classes: Strategies for Effective Course Management},
 booktitle = {Proceedings of the Twenty-ninth SIGCSE Technical Symposium on Computer Science Education},
 series = {SIGCSE '98},
 year = {1998},
 isbn = {0-89791-994-7},
 pages = {131--134},
 numpages = {4},
 url = {http://doi.acm.org/10.1145/273133.273177},
 doi = {10.1145/273133.273177},
 acmid = {273177},
 publisher = {ACM},
 address = {New York, NY, USA},
}

@inproceedings{Reges:1988,
 author = {Reges, Stuart and McGrory, John and Smith, Jeff},
 title = {The Effective Use of Undergraduates to Staff Large Introductory CS Courses},
 booktitle = {Proceedings of the Nineteenth SIGCSE Technical Symposium on Computer Science Education},
 series = {SIGCSE '88},
 year = {1988},
 isbn = {0-89791-256-X},
 location = {Atlanta, Georgia, USA},
 pages = {22--25},
 numpages = {4},
 url = {http://doi.acm.org/10.1145/52964.52971},
 doi = {10.1145/52964.52971},
 acmid = {52971},
 publisher = {ACM},
 address = {New York, NY, USA},
}

@inproceedings{Roberts:1995,
 author = {Roberts, Eric and Lilly, John and Rollins, Bryan},
 title = {Using Undergraduates As Teaching Assistants in Introductory Programming Courses: An Update on the Stanford Experience},
 booktitle = {Proceedings of the Twenty-sixth SIGCSE Technical Symposium on Computer Science Education},
 series = {SIGCSE '95},
 year = {1995},
 isbn = {0-89791-693-X},
 pages = {48--52},
 numpages = {5},
 url = {http://doi.acm.org/10.1145/199688.199716},
 doi = {10.1145/199688.199716},
 acmid = {199716},
 publisher = {ACM},
 address = {New York, NY, USA},
}

@inproceedings{Reges:2003,
 author = {Reges, Stuart},
 title = {Using Undergraduates As Teaching Assistants at a State University},
 booktitle = {Proceedings of the 34th SIGCSE Technical Symposium on Computer Science Education},
 series = {SIGCSE '03},
 year = {2003},
 isbn = {1-58113-648-X},
 pages = {103--107},
 numpages = {5},
 url = {http://doi.acm.org/10.1145/611892.611943},
 doi = {10.1145/611892.611943},
 acmid = {611943},
 publisher = {ACM},
 address = {New York, NY, USA},
}

@inproceedings{Dickson:2017,
 author = {Dickson, Paul E. and Dragon, Toby and Lee, Adam},
 title = {Using Undergraduate Teaching Assistants in Small Classes},
 booktitle = {Proceedings of the 2017 ACM SIGCSE Technical Symposium on Computer Science Education},
 series = {SIGCSE '17},
 year = {2017},
 isbn = {978-1-4503-4698-6},
 pages = {165--170},
 numpages = {6},
 url = {http://doi.acm.org/10.1145/3017680.3017725},
 doi = {10.1145/3017680.3017725},
 acmid = {3017725},
 publisher = {ACM},
 address = {New York, NY, USA},
}

@inproceedings{Decker:2006,
 author = {Decker, Adrienne and Ventura, Phil and Egert, Christopher},
 title = {Through the Looking Glass: Reflections on Using Undergraduate Teaching Assistants in CS1},
 booktitle = {Proceedings of the 37th SIGCSE Technical Symposium on Computer Science Education},
 series = {SIGCSE '06},
 year = {2006},
 isbn = {1-59593-259-3},
 pages = {46--50},
 numpages = {5},
 url = {http://doi.acm.org/10.1145/1121341.1121358},
 doi = {10.1145/1121341.1121358},
 acmid = {1121358},
 publisher = {ACM},
 address = {New York, NY, USA},
}

@article{Lewis:2017,
 author = {Lewis, Colleen M.},
 title = {ACM Retention Committee: Twelve Tips for Creating a Culture That Supports All Students in Computing},
 journal = {ACM Inroads},
 issue_date = {December 2017},
 volume = {8},
 number = {4},
 month = oct,
 year = {2017},
 issn = {2153-2184},
 pages = {17--20},
 numpages = {4},
 url = {http://doi.acm.org/10.1145/3148524},
 doi = {10.1145/3148524},
 acmid = {3148524},
 publisher = {ACM},
 address = {New York, NY, USA},
}

@report{RetentionCS,
 author = {Stephenson, Chris and Derbenwick Miller, Alison and Alvarado, Christine and Barker, Lecia and Barr, Valerie and Camp, Tracy and Frieze, Carol and Lewis, Colleen and Cannon Mindell, Erin and Limbird, Lee and Richardson, Debra and Sahami, Mehran and Villa, Elsa and Walker, Henry and Zweben, Stuart},
 title = {Retention in Computer Science Undergraduate Programs in the U.S.: Data Challenges and Promising Interventions},
 year = {2018},
 url = {https://www.acm.org/binaries/content/assets/education/retention-in-cs-undergrad-programs-in-the-us.pdf},
 publisher = {ACM},
 address = {New York, NY, USA},
}

@inproceedings{Patitsas:2014,
 author = {Patitsas, Elizabeth and Craig, Michelle and Easterbrook, Steve},
 title = {A Historical Examination of the Social Factors Affecting Female Participation in Computing},
 booktitle = {Proceedings of the 2014 Conference on Innovation \& Technology in Computer Science Education},
 series = {ITiCSE '14},
 year = {2014},
 isbn = {978-1-4503-2833-3},
 location = {Uppsala, Sweden},
 pages = {111--116},
 numpages = {6},
 url = {http://doi.acm.org/10.1145/2591708.2591731},
 doi = {10.1145/2591708.2591731},
 acmid = {2591731},
 publisher = {ACM},
 address = {New York, NY, USA},
}

@inproceedings{Patitsas:2016, 
 author = {Patitsas, Elizabeth and Craig, Michelle and Easterbrook, Steve}, 
 title = {How CS departments are managing the enrolment boom: Troubling implications for diversity}, 
 booktitle = {2016 Research on Equity and Sustained Participation in Engineering, Computing, and Technology (RESPECT)}, 
 year = {2016}, 
 pages = {1-2}, 
 doi = {10.1109/RESPECT.2016.7836180}, 
 month = aug, 
}

@inproceedings{Smith:2017,
 author = {Smith, Aaron J. and Boyer, Kristy Elizabeth and Forbes, Jeffrey and Heckman, Sarah and Mayer-Patel, Ketan},
 title = {My Digital Hand: A Tool for Scaling Up One-to-One Peer Teaching in Support of Computer Science Learning},
 booktitle = {Proceedings of the 2017 ACM SIGCSE Technical Symposium on Computer Science Education},
 series = {SIGCSE '17},
 year = {2017},
 isbn = {978-1-4503-4698-6},
 location = {Seattle, Washington, USA},
 pages = {549--554},
 numpages = {6},
 url = {http://doi.acm.org/10.1145/3017680.3017800},
 doi = {10.1145/3017680.3017800},
 acmid = {3017800},
 publisher = {ACM},
 address = {New York, NY, USA},
}

@article{MacWilliam:2013,
 author = {MacWilliam, Tommy and Malan, David J.},
 title = {Scaling Office Hours: Managing Live Q\&\#38;A in Large Courses},
 journal = {J. Comput. Sci. Coll.},
 issue_date = {January 2013},
 volume = {28},
 number = {3},
 month = jan,
 year = {2013},
 issn = {1937-4771},
 pages = {94--101},
 numpages = {8},
 url = {http://dl.acm.org/citation.cfm?id=2400161.2400179},
 acmid = {2400179},
 publisher = {Consortium for Computing Sciences in Colleges},
 address = {USA},
}

@inproceedings{Basu:2015,
 author = {Basu, Soumya and Wu, Albert and Hou, Brian and DeNero, John},
 title = {Problems Before Solutions: Automated Problem Clarification at Scale},
 booktitle = {Proceedings of the Second (2015) ACM Conference on Learning @ Scale},
 series = {L@S '15},
 year = {2015},
 isbn = {978-1-4503-3411-2},
 location = {Vancouver, BC, Canada},
 pages = {205--213},
 numpages = {9},
 url = {http://doi.acm.org/10.1145/2724660.2724679},
 doi = {10.1145/2724660.2724679},
 acmid = {2724679},
 publisher = {ACM},
 address = {New York, NY, USA},
}

@inproceedings{Sridhara:2016,
 author = {Sridhara, Sumukh and Hou, Brian and Lu, Jeffrey and DeNero, John},
 title = {Fuzz Testing Projects in Massive Courses},
 booktitle = {Proceedings of the Third (2016) ACM Conference on Learning @ Scale},
 series = {L@S '16},
 year = {2016},
 isbn = {978-1-4503-3726-7},
 location = {Edinburgh, Scotland, UK},
 pages = {361--367},
 numpages = {7},
 url = {http://doi.acm.org/10.1145/2876034.2876050},
 doi = {10.1145/2876034.2876050},
 acmid = {2876050},
 publisher = {ACM},
 address = {New York, NY, USA},
}

@inproceedings{Phothilimthana:2017,
 author = {Phothilimthana, Phitchaya Mangpo and Sridhara, Sumukh},
 title = {High-Coverage Hint Generation for Massive Courses: Do Automated Hints Help CS1 Students?},
 booktitle = {Proceedings of the 2017 ACM Conference on Innovation and Technology in Computer Science Education},
 series = {ITiCSE '17},
 year = {2017},
 isbn = {978-1-4503-4704-4},
 location = {Bologna, Italy},
 pages = {182--187},
 numpages = {6},
 url = {http://doi.acm.org/10.1145/3059009.3059058},
 doi = {10.1145/3059009.3059058},
 acmid = {3059058},
 publisher = {ACM},
 address = {New York, NY, USA},
}

@inproceedings{Yan:2018,
 author = {Yan, Lisa and McKeown, Nick and Sahami, Mehran and Piech, Chris},
 title = {TMOSS: Using Intermediate Assignment Work to Understand Excessive Collaboration in Large Classes},
 booktitle = {Proceedings of the 49th ACM Technical Symposium on Computer Science Education},
 series = {SIGCSE '18},
 year = {2018},
 isbn = {978-1-4503-5103-4},
 location = {Baltimore, Maryland, USA},
 pages = {110--115},
 numpages = {6},
 url = {http://doi.acm.org/10.1145/3159450.3159490},
 doi = {10.1145/3159450.3159490},
 acmid = {3159490},
 publisher = {ACM},
 address = {New York, NY, USA},
}

@inproceedings{Yan:2019,
 author = {Yan, Lisa and Hu, Annie and Piech, Chris},
 title = {Pensieve: Feedback on Coding Process for Novices},
 booktitle = {Proceedings of the 50th ACM Technical Symposium on Computer Science Education},
 series = {SIGCSE '19},
 year = {2019},
 isbn = {978-1-4503-5890-3},
 location = {Minneapolis, MN, USA},
 pages = {253--259},
 numpages = {7},
 url = {http://doi.acm.org/10.1145/3287324.3287483},
 doi = {10.1145/3287324.3287483},
 acmid = {3287483},
 publisher = {ACM},
 address = {New York, NY, USA},
}

@inproceedings{Piech:2012,
 author = {Piech, Chris and Sahami, Mehran and Koller, Daphne and Cooper, Steve and Blikstein, Paulo},
 title = {Modeling How Students Learn to Program},
 booktitle = {Proceedings of the 43rd ACM Technical Symposium on Computer Science Education},
 series = {SIGCSE '12},
 year = {2012},
 isbn = {978-1-4503-1098-7},
 location = {Raleigh, North Carolina, USA},
 pages = {153--160},
 numpages = {8},
 url = {http://doi.acm.org/10.1145/2157136.2157182},
 doi = {10.1145/2157136.2157182},
 acmid = {2157182},
 publisher = {ACM},
 address = {New York, NY, USA},
}

@inproceedings{Wang:2017,
 author = {Wang, Lisa and Sy, Angela and Liu, Larry and Piech, Chris},
 title = {Deep Knowledge Tracing On Programming Exercises},
 booktitle = {Proceedings of the Fourth (2017) ACM Conference on Learning @ Scale},
 series = {L@S '17},
 year = {2017},
 isbn = {978-1-4503-4450-0},
 location = {Cambridge, Massachusetts, USA},
 pages = {201--204},
 numpages = {4},
 url = {http://doi.acm.org/10.1145/3051457.3053985},
 doi = {10.1145/3051457.3053985},
 acmid = {3053985},
 publisher = {ACM},
 address = {New York, NY, USA},
}

@inproceedings{Liao:2019,
 author = {Liao, Soohyun Nam and Zingaro, Daniel and Alvarado, Christine and Griswold, William G. and Porter, Leo},
 title = {Exploring the Value of Different Data Sources for Predicting Student Performance in Multiple CS Courses},
 booktitle = {Proceedings of the 50th ACM Technical Symposium on Computer Science Education},
 series = {SIGCSE '19},
 year = {2019},
 isbn = {978-1-4503-5890-3},
 location = {Minneapolis, MN, USA},
 pages = {112--118},
 numpages = {7},
 url = {http://doi.acm.org/10.1145/3287324.3287407},
 doi = {10.1145/3287324.3287407},
 acmid = {3287407},
 publisher = {ACM},
 address = {New York, NY, USA},
}

@article{Glassman:2015,
 author = {Glassman, Elena L. and Scott, Jeremy and Singh, Rishabh and Guo, Philip J. and Miller, Robert C.},
 title = {OverCode: Visualizing Variation in Student Solutions to Programming Problems at Scale},
 journal = {ACM Trans. Comput.-Hum. Interact.},
 issue_date = {April 2015},
 volume = {22},
 number = {2},
 month = mar,
 year = {2015},
 issn = {1073-0516},
 pages = {7:1--7:35},
 articleno = {7},
 numpages = {35},
 url = {http://doi.acm.org/10.1145/2699751},
 doi = {10.1145/2699751},
 acmid = {2699751},
 publisher = {ACM},
 address = {New York, NY, USA},
}

@inproceedings{Astrachan:2007,
 author = {Astrachan, Owen and Parlante, Nick and Garcia, Daniel D. and Reges, Stuart},
 title = {Teaching Tips We Wish They'd Told Us Before We Started},
 booktitle = {Proceedings of the 38th SIGCSE Technical Symposium on Computer Science Education},
 series = {SIGCSE '07},
 year = {2007},
 isbn = {1-59593-361-1},
 location = {Covington, Kentucky, USA},
 pages = {2--3},
 numpages = {2},
 url = {http://doi.acm.org/10.1145/1227310.1227314},
 doi = {10.1145/1227310.1227314},
 acmid = {1227314},
 publisher = {ACM},
 address = {New York, NY, USA},
}

@article{Titterton:2010,
 author = {Titterton, Nathaniel and Lewis, Colleen M. and Clancy, Michael J.},
 title = {Experiences with lab-centric instruction},
 journal = {Computer Science Education},
 volume = {20},
 number = {2},
 pages = {79-102},
 year  = {2010},
 publisher = {Routledge},
 doi = {10.1080/08993408.2010.486256},
 url = {https://doi.org/10.1080/08993408.2010.486256},
 eprint = {https://doi.org/10.1080/08993408.2010.486256}
}

@inproceedings{West:2016,
 author = {West, Matthew and Zilles, Craig},
 title = {Modeling Student Scheduling Preferences in a Computer-Based Testing Facility},
 booktitle = {Proceedings of the Third (2016) ACM Conference on Learning @ Scale},
 series = {L@S '16},
 year = {2016},
 isbn = {978-1-4503-3726-7},
 location = {Edinburgh, Scotland, UK},
 pages = {309--312},
 numpages = {4},
 url = {http://doi.acm.org/10.1145/2876034.2893441},
 doi = {10.1145/2876034.2893441},
 acmid = {2893441},
 publisher = {ACM},
 address = {New York, NY, USA},
}

@inproceedings{Nip:2018,
 author = {Nip, Terence and Gunter, Elsa L. and Herman, Geoffrey L. and Morphew, Jason W. and West, Matthew},
 title = {Using a Computer-based Testing Facility to Improve Student Learning in a Programming Languages and Compilers Course},
 booktitle = {Proceedings of the 49th ACM Technical Symposium on Computer Science Education},
 series = {SIGCSE '18},
 year = {2018},
 isbn = {978-1-4503-5103-4},
 location = {Baltimore, Maryland, USA},
 pages = {568--573},
 numpages = {6},
 url = {http://doi.acm.org/10.1145/3159450.3159500},
 doi = {10.1145/3159450.3159500},
 acmid = {3159500},
 publisher = {ACM},
 address = {New York, NY, USA},
}

@inproceedings{Piech:2018,
 author = {Piech, Chris and Gregg, Chris},
 title = {BlueBook: A Computerized Replacement for Paper Tests in Computer Science},
 booktitle = {Proceedings of the 49th ACM Technical Symposium on Computer Science Education},
 series = {SIGCSE '18},
 year = {2018},
 isbn = {978-1-4503-5103-4},
 location = {Baltimore, Maryland, USA},
 pages = {562--567},
 numpages = {6},
 url = {http://doi.acm.org/10.1145/3159450.3159587},
 doi = {10.1145/3159450.3159587},
 acmid = {3159587},
 publisher = {ACM},
 address = {New York, NY, USA},
}

@inproceedings{Pon-Barry:2019,
 author = {Pon-Barry, Heather and St. John, Audrey and Packard, Becky Wai-Ling and Rotundo, Barbara},
 title = {A Flexible Curriculum for Promoting Inclusion Through Peer Mentorship},
 booktitle = {Proceedings of the 50th ACM Technical Symposium on Computer Science Education},
 series = {SIGCSE '19},
 year = {2019},
 isbn = {978-1-4503-5890-3},
 location = {Minneapolis, MN, USA},
 pages = {1116--1122},
 numpages = {7},
 url = {http://doi.acm.org/10.1145/3287324.3287434},
 doi = {10.1145/3287324.3287434},
 acmid = {3287434},
 publisher = {ACM},
 address = {New York, NY, USA},
}

@inproceedings{Prather:2019,
 author = {Prather, James and Pettit, Raymond and Becker, Brett A. and Denny, Paul and Loksa, Dastyni and Peters, Alani and Albrecht, Zachary and Masci, Krista},
 title = {First Things First: Providing Metacognitive Scaffolding for Interpreting Problem Prompts},
 booktitle = {Proceedings of the 50th ACM Technical Symposium on Computer Science Education},
 series = {SIGCSE '19},
 year = {2019},
 isbn = {978-1-4503-5890-3},
 location = {Minneapolis, MN, USA},
 pages = {531--537},
 numpages = {7},
 url = {http://doi.acm.org/10.1145/3287324.3287374},
 doi = {10.1145/3287324.3287374},
 acmid = {3287374},
 publisher = {ACM},
 address = {New York, NY, USA},
}

\end{document}